\documentclass[aps,prl,twocolumn,superscriptaddress,groupedaddress,showpacs,nofootinbib]{revtex4}

\usepackage[english]{babel}
\usepackage{amsmath}
\usepackage{amsthm}
\usepackage{amssymb}
\usepackage{booktabs}
\usepackage{color}
\usepackage{hyperref}
\hypersetup{
  colorlinks   = true,  
  urlcolor     = blue,  
  linkcolor    = blue,  
  citecolor    = red    
}
\usepackage{natbib}
\usepackage{pgfplots}
\usepackage{subfigure}
\usepackage{url}
\newcommand{\ket}[1]{ | #1 \rangle }

\newcommand{\braket}[2]{ \langle #1 | #2 \rangle }

\newcommand{\refFigure}[1]{{\textrm{Fig.~\ref{#1}}}}

\newcommand{\refEquation}[1]{{\textrm{Eq.~(\ref{#1})}}}

\begin{document}

\title{Trapped electrons in the quantum degenerate regime}

\author{Ronen \surname{Kroeze}}
\email{r.m.kroeze@student.tue.nl}

\author{Jom \surname{Luiten}}
\affiliation{Eindhoven University of Technology, P.~O.~Box 513, 5600 MB Eindhoven, The Netherlands}

\author{Servaas \surname{Kokkelmans}}
\affiliation{Eindhoven University of Technology, P.~O.~Box 513, 5600 MB Eindhoven, The Netherlands}

\date{\today}


\begin{abstract}
A full strength Coulomb interaction between trapped electrons can be felt only in absence of a neutralizing background. In order to study quantum degenerate electrons without such a background, an external trap is needed to compensate for the strong electronic repulsion.
As a basic model for such a system, we study a trapped electron pair in a harmonic trap with an explicit inclusion of its Coulomb interaction. We find the eigenenergy of the ground state, confirming earlier work in the context of harmonium. We extend this to a complete set of properly scaled energies for any value of the trapping strength, including the excited states. The problem is solved either numerically or by making harmonic approximations to the potential. As function of the trapping strength a crossover can be made from the strongly to the weakly-coupled regime, and we show that in both regimes perturbative methods based on a pair-wise electron description would be effective for a many-particle trapped electron system, which resembles a Wigner crystal in the ground state of the strongly coupled limit.
\end{abstract}

\maketitle

\section{Introduction}
The strong Coulomb interaction between electrons is rarely felt at full strength, as their repulsion is commonly screened by the presence of positive charges. This is usual fashion in condensed matter physics, where the valence electrons can be treated as a weakly-interacting electron gas. Such systems are said to be in the weakly to intermediate coupling regime, which has been successfully described by Landau Fermi Liquid Theory. This weakly interacting electron gas is also quantum degenerate in most metals at room temperature.

The coupling parameter of these many-particle systems is generally considered as the ratio of potential energy over kinetic energy. Also plasma physics systems are generally weakly coupled: here screening is governed by the Debye length, which make the effective electron interactions of short range nature. Only in charged particle accelerators, the pure Coulomb interaction can be appreciated at full strength. However at the same time, this strong repulsion makes it very difficult for an electron gas to become strongly-coupled as a result of heating, which effectively prevents quantum degeneracy as well.

Nevertheless, there is an ever higher demand for brightness in the field of charged particle accelerators. Here the holy grail would be to achieve ultimate conditions of brightness, by creating quantum degenerate electron or ion beams. Particle accelerators are heavily used in medical and industrial applications and for driving high-brightness X-ray light sources \cite{AcceleratorsAndBeams}. Ultimately this would require electron beam phase space densities close to degeneracy, three to four orders of magnitude higher than in present state-of-the-art electron microscopy beams \cite{luiten2007ultracold}. In this paper, we investigate some basic properties of trapped electrons, which are well-behaved systems where heating effects can be controlled, and we investigate different regimes of coupling and degeneracy. The trapping frequency $\omega$ acts as a control parameter, which allows us to study a crossover between different limiting coupling regimes. One limit is a strongly-coupled freely propagating electron gas, which could be an electron beam for instance. In this regime a Wigner crystal is known to exist as the ground state \cite{wigner1934interaction}. The other limit, for large trapping frequencies, is then a weakly-coupled trapped electron gas.

Our approach allows for new opportunities in perturbatively describing few- and many-particle Coulomb systems. We offer a simple description of the crossover in terms of a reference frequency, that depends on the electron mass. We extend to other types of charged particles including different masses, and discuss the feasibility to investigate the crossover in these systems, based on mass and trapping possibilities.

Cold atomic gases offer a promising route towards ultimate brightness conditions. In particular Rydberg crystals are very close to the quantum plasma regime, and extraction of the electrons could be done with a minimum amount of heating, such that the electrons could be trapped in Paul traps \cite{paul_trap}, Penning traps \cite{gabrielse_review,gabrielse_two_electron}, and Ponderomotive traps \cite{ponderomotive_trap}. Also other ways towards degeneracy are currently being explored, such as  hybrid field electron extraction using femtosecond laser pulses \cite{Lougovski11}, direct cooling in ion traps, and sympathetic cooling with atoms \cite{Balewski13,Schmid12}.

The quantum-mechanical problem of particles with short-range interactions in a harmonic oscillator has been well studied \cite{Busch,Mentink}. Moreover, appropriate studies where the interaction is of the Coulomb type have been done as well \cite{SamGosh,Taut,kestner1962study}. Identical particles in such a trap are referred to as {\it harmonium} \cite{CioPer,Karwowski04,Cioslowski12,Cioslowski06,CioslowskiPRA}, which has been studied as an exactly solvable model for the artificial helium atom, where the interaction of the electrons with the nucleus has been replaced by a harmonic confinement.  Analytical solutions have been found for a certain set of oscillator frequencies \cite{Taut}, numerical solutions that interpolate between them \cite{Karwowski04}, as well as an analytical interpolation model \cite{Cioslowski12}. Here we demonstrate another set of analytical interpolating solutions, and we show that the effective Coulomb interaction between the particles can in certain cases be considered as a short-range interaction.

The problem is approached by writing the wave function in the angular momentum basis. In the problem of Coulomb scattering this basis is not convenient since the phase shift does not converge, in other words, the Coulomb scattering problem is never asymptotically free \cite{sakurai}. In our problem, however, the harmonic trap cures this problem; the eigenstates of the harmonic oscillator take over the role of the asymptotic free states. Therefore, the angular momentum basis is indeed convenient. As a result, the eigenstates of the harmonic oscillator define our reference energy which will be taken as $\hbar\omega$.

\section{Problem description}
When considering two trapped electrons, it is convenient to transform to the center of mass and relative coordinates. With this transformation, the problem in the center of mass reduces to a three dimensional harmonic oscillator, which is easily solved \cite{bransden2000quantum}, whereas the motion in the relative coordinate $\mathbf{r}$ is governed by a three dimensional Schr\"odinger equation
\begin{equation}
\hat{H}\ket{\Psi}\equiv\left(\frac{\hat{p}^2}{2\mu}+V(r)\right)\ket{\Psi}=E\ket{\Psi}
\end{equation}
where $\mu=\frac{1}{2}m_e$ is the effective mass and the potential
\begin{equation}\label{Total_pot}
V(r)=\frac{1}{2}\mu\omega^2r^2+\frac{e^2}{4\pi\varepsilon_0 r}
\end{equation}
only depends on the length of the relative coordinate $r=|\mathbf{r}|$. $\omega$ is the frequency of the harmonic oscillator and the second term is the Coulomb interaction for two electrons. Since \refEquation{Total_pot} is a central potential we can reduce this further to a one-dimensional Schr\"odinger equation which reads
\begin{equation}\label{Schrod_princ}
-\frac{\hbar^2}{2\mu}\frac{d^2u}{dr^2}+\left(\frac{1}{2}\mu\omega^2r^2+ \frac{e^2}{4\pi\varepsilon_0 r}+\frac{\hbar^2l(l+1)}{2\mu r^2}\right)u=Eu,
\end{equation}
by writing the wave function in the angular momentum basis
\begin{equation}
\braket{\mathbf r}{\Psi}=\frac{u(r)}{r}Y_{lm}(\hat{\mathbf{r}})
\end{equation}
with quantum numbers $l$ and $m$ as usual. Thus the problem is reduced from solving a six-dimensional eigenproblem to solving the one-dimensional eigenproblem given by \refEquation{Schrod_princ}. As mentioned before, the angular momentum basis is a good basis since the harmonic oscillator states serve as the asymptotic free states, which do not exist for a simple Coulomb scattering problem.

\section{Exact solutions}
In general, the Schr\"odinger equation \refEquation{Schrod_princ} is not analytically solvable, but rather the problem is quasi-exactly solvable \cite{quasi_exact_1,quasi_exact_2}. This problem has been studied by Taut \cite{Taut}, amongst others. Taut has shown that exact solutions can be found by assuming that the wave functions can be written as
\begin{equation}
u(r)=P(\rho)\exp\left(-\frac{1}{2}\rho^2\right),
\end{equation}
where $\rho=\sqrt{\frac{\mu\omega}{\hbar}}r$ and
\begin{equation}
P(\rho)=\rho^{l+1}\sum\limits_{i=0}^j a_i \rho^i
\end{equation}
is a polynomial of finite degree. With this Ansatz, \refEquation{Schrod_princ} yields a recurrence relation for the coefficients $a_i$, which can be solved when $j$ is given. The corresponding solutions are shown in \refFigure{Num_vs_Taut}, where we show the eigenenergies normalized with respect to the oscillator energy $\hbar\omega$ versus the oscillator frequency $\omega$. Discussion of the eigenenergies is delayed to the next section where full solutions can be found. Taut's solutions correspond only to a countable number of solutions (namely integer $j$) where no solution can be found above a frequency of $\frac{m_e e^4}{32\pi^2\varepsilon_0^2 \hbar^3}=\frac{1}{2}\omega_0$, which is, as we will see below, half the Hartree frequency. As we seek solutions for any set of parameters we thus need other methods.

\section{Full solutions}
\begin{figure}[!hbt]
\begin{center}
\includegraphics[width=0.45\textwidth]{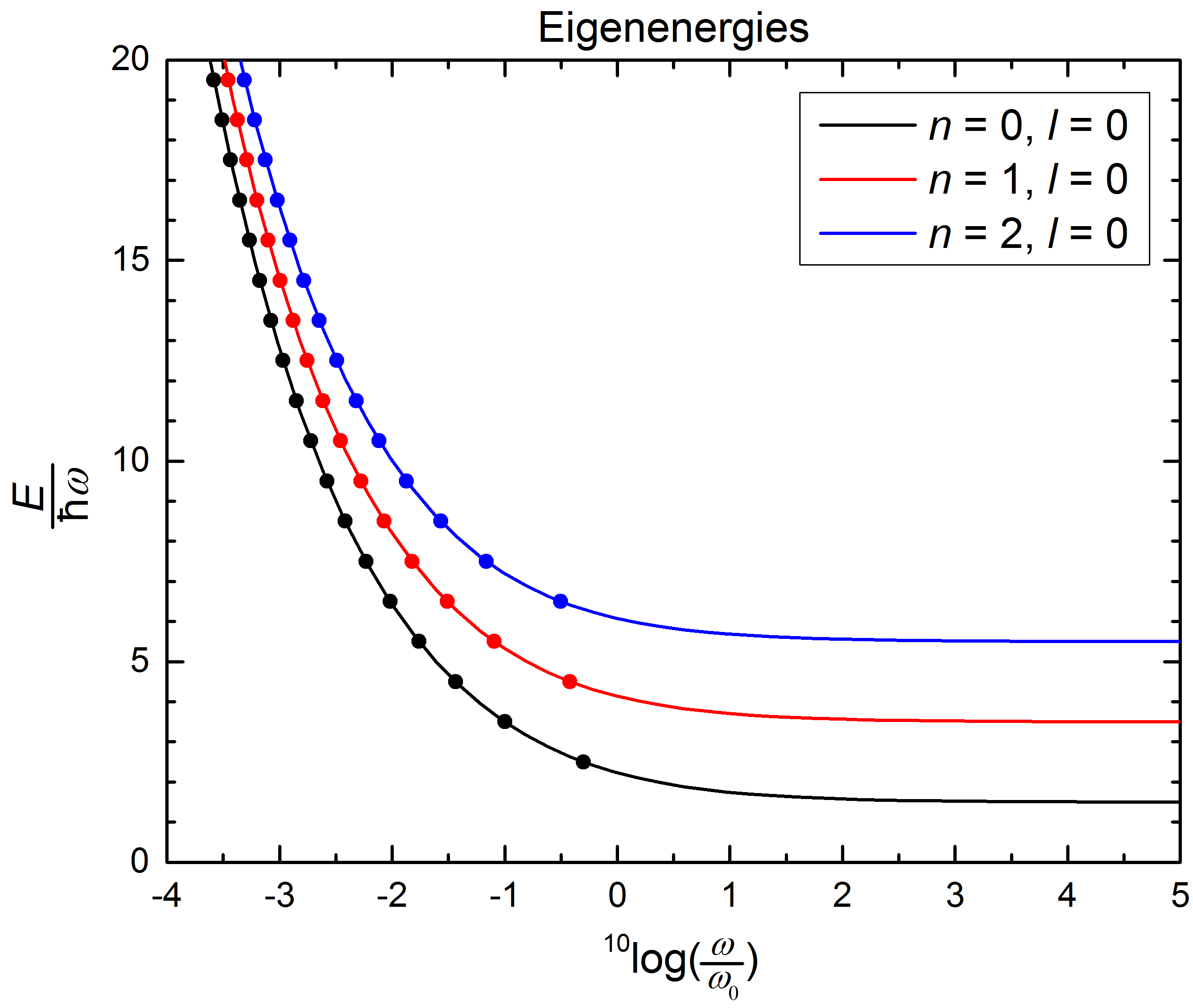}
\end{center}
\vspace{-0.75cm}
\caption{\textrm{Exact eigenenergies as found by Taut's method (dots) as well as the numerical solutions (lines) of the differential equation \refEquation{Schrod_princ} as function of the frequency $\omega$. $\omega_0$ is a typical frequency of the problem and is exactly two times the highest frequency for which Taut's method can find an exact solution.}}
\label{Num_vs_Taut}
\end{figure}

A complete solution of the problem, i.e. for every set of parameters, can be found by solving \refEquation{Schrod_princ} numerically. The eigenenergies are shown as solid lines in \refFigure{Num_vs_Taut} and match Taut's solutions as expected. We observe two distinct regimes. First, for large frequencies we see that the energies are approximately that of a three dimensional harmonic oscillator, or in the limit of $\omega\rightarrow\infty$ these seem to match exactly. We thus can conclude that in the regime of large frequencies the Coulomb repulsion is ineffective and we call this the \textit{free harmonic limit}. Accordingly, for large frequencies the Coulomb interaction can be considered for this system as a short range interaction. This is a counter-intuitive result, as a large frequency corresponds to a small separation distance $r$, and therefore to a large Coulomb interaction energy.

On the other hand, when the frequency is small we see that the energy diverges from that of a harmonic oscillator. Using the definition of the weakly and strongly coupled regimes we can identify this regime as the strongly coupled / strongly interacting regime. In this regime the Coulomb repulsion is relatively strong. We can relate both frequency regimes to densities since a smaller frequency corresponds to a smaller separation distance $r$ and thus a lower density. So concluding, for a low density the interaction is strong and for a high density the interaction is weak. This is exactly opposite to what one would expect for a short-range interaction force in an external field.

\section{Harmonic approximations}
In this section, we derive analytical solutions for this problem by using an approximate expression for the effective radial potential
\begin{equation}\label{Effective_pot}
V_{\text{eff}}(r)=\frac{1}{2}\mu\omega^2r^2+\frac{e^2}{4\pi\varepsilon_0 r}+\frac{\hbar^2l(l+1)}{2\mu r^2}.
\end{equation}
This was first done by Taut \cite{Taut} and in the context of harmonium \cite{CioPer,Karwowski04} for a partial list of energies, however, here we extend this to a complete set of properly scaled energies, including the excited states.

In the free harmonic limit we use
\begin{equation}
\tilde{V}(r)=\frac{1}{2}\mu\omega^2r^2
\end{equation}
to find
\begin{equation}\label{Energy_high_w}
\frac{E}{\hbar\omega}=2n+l+\frac{3}{2},
\end{equation}
which indeed corresponds to the 3D harmonic oscillator energies found for large $\omega$ as mentioned in the previous section. For small frequencies, thus in the strongly interacting regime, we make a Taylor approximation of the potential around its minimum:
\begin{equation}\label{HO_low_w}
\tilde{V}(r)=V_0+\frac{1}{2}\mu\tilde{\omega}^2\left(r-r_0\right)^2.
\end{equation}
For $l=0$ we find the eigenenergies
\begin{equation}\label{Energy_low_w}
\frac{E_{l=0}}{\hbar\omega}=\frac{3}{2^{4/3}}\left(\frac{\omega}{\omega_0}\right)^{-1/3}+\left(n+\frac{1}{2}\right)\sqrt{3}.
\end{equation}
where $\omega_0=\frac{m_e e^4}{16\pi^2\varepsilon_0^2 \hbar^3}=E_h/\hbar$ is the Hartree frequency, which is the typical frequency for this problem. It corresponds to the situation where the minimum of \refEquation{Effective_pot} is exactly $\sqrt[3]{2}$ times the Bohr radius, $r_0=\sqrt[3]{2}a_0\equiv \frac{2^{7/3}\pi\varepsilon_0\hbar^2}{m_e e^2}$.
The first term corresponds to the divergence of $\frac{E}{\hbar\omega}$ for small frequencies given by a typical exponent -1/3 and a proportionality constant directly related to $\omega_0$. For non-zero $l$ using the same approximation we recover the same asymptotic behavior for small frequencies.

\begin{figure}[!hbt]
\begin{center}
\includegraphics[width=0.45\textwidth]{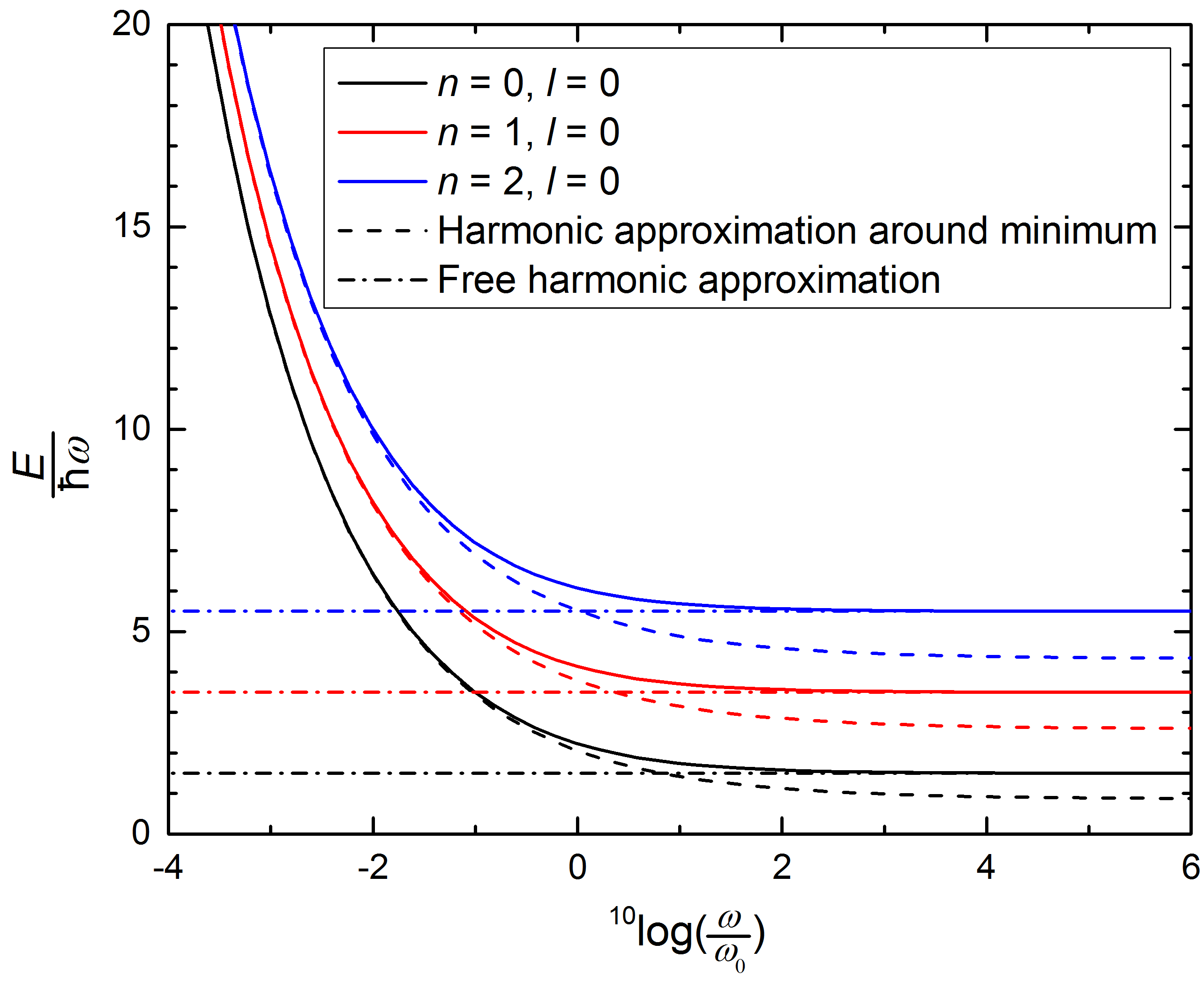}
\end{center}
\vspace{-0.75cm}
\caption{\textrm{Comparison of numerical eigenenergies (solid) with the eigenenergies \refEquation{Energy_high_w} from the free harmonic approximations (dash-dot) and the eigenenergies \refEquation{Energy_low_w} from the harmonic approximation around the minimum (dashed).}}
\label{2HO_vs_Numerical}
\vspace{-1em}
\end{figure}

The second term in \refEquation{Energy_low_w} corresponds to the spacing between different levels. The approximate eigenenergies given by \refEquation{Energy_high_w} and \refEquation{Energy_low_w} can be compared to the numerical eigenenergies, see \refFigure{2HO_vs_Numerical}. From \refFigure{2HO_vs_Numerical} we see that the harmonic approximations work well in both the strongly and weakly interacting regime.

\section{Hellmann-Feynman}
Using the Hellmann-Feynman theorem \cite{H,F}, given here as
\begin{equation}
\frac{dE}{d\omega}=\mu\omega\int\limits_0^\infty r^2\left|u(r)\right|^2dr=\mu\omega\langle r^2\rangle,
\end{equation}
we are able to calculate $\frac{dE}{d\omega}$ when we know the solution $u(r)$, which gives us additional information about the eigenenergies we are looking for. We assume $u(r)$ is a flat wave function on the classically allowed domain of the harmonic oscillator potential. This is justified at large relative energies and a sufficiently low frequency, which corresponds to the strongly interacting regime. With this approximation we can determine $\langle r^2\rangle$ for $n=l=0$ and find
\begin{equation}\label{HF_diff}
\frac{dE}{d\omega}=\frac{2E}{3\omega},
\end{equation}
which is a differential equation for $E(\omega)$. The solution is
\begin{equation}
E(\omega)=A\omega^{2/3},
\label{scalinglaw}
\end{equation}
which is exactly the same asymptotic behavior as given by equation \refEquation{Energy_low_w}. In fact, the proportionality constant $A$ can be found by approximating $\langle r^2\rangle\approx r_0^2$, where $r_0$ is the minimum of the effective potential. Then, $A=\frac{3}{2^{4/3}}\omega_0^{1/3}$ is found, consistent with \refEquation{Energy_low_w}.

Moreover, with the correct asymptote for $\omega\rightarrow\infty$ given by \refEquation{Energy_high_w} and with a suitable two-parameter interpolation function we find a more complete differential equation for general $n$ and $l$, namely
\begin{equation}\label{HF_diff_better}
\begin{split}
\frac{dE}{d\omega}=&\frac{2E}{3\omega}+\frac{\hbar}{3}\left(2n+l+\frac{3}{2}\right)\\
&+\frac{\hbar}{3}\frac{1}{1+\alpha\omega^\beta}\left[\left(n+\frac{1}{2}\right)\sqrt{3}-\left(2n+l+\frac{3}{2}\right)\right],
\end{split}
\end{equation}
where $\alpha$ and $\beta$ are fitting parameters. The frequency dependence of the last term is an empirical choice that interpolates between Taut's solutions, for which $\langle r^2\rangle$ can be calculated analytically. After fitting through the exact solutions, the above differential equation can be solved analytically to find the eigenenergy as function of frequency. Note that this provides us with a fundamentally different interpolation scheme than the one in \cite{Cioslowski12}; whereas there the Hellman-Feynman theorem is used after establishing $E(\omega)$ here it is used predictively. Another significant difference is that our method easily allows for interpolation of excited states ($n\neq0$ or $l\neq0$).

\section{General states}
\begin{figure}[!hbt]
\begin{center}
\includegraphics[width=0.48\textwidth]{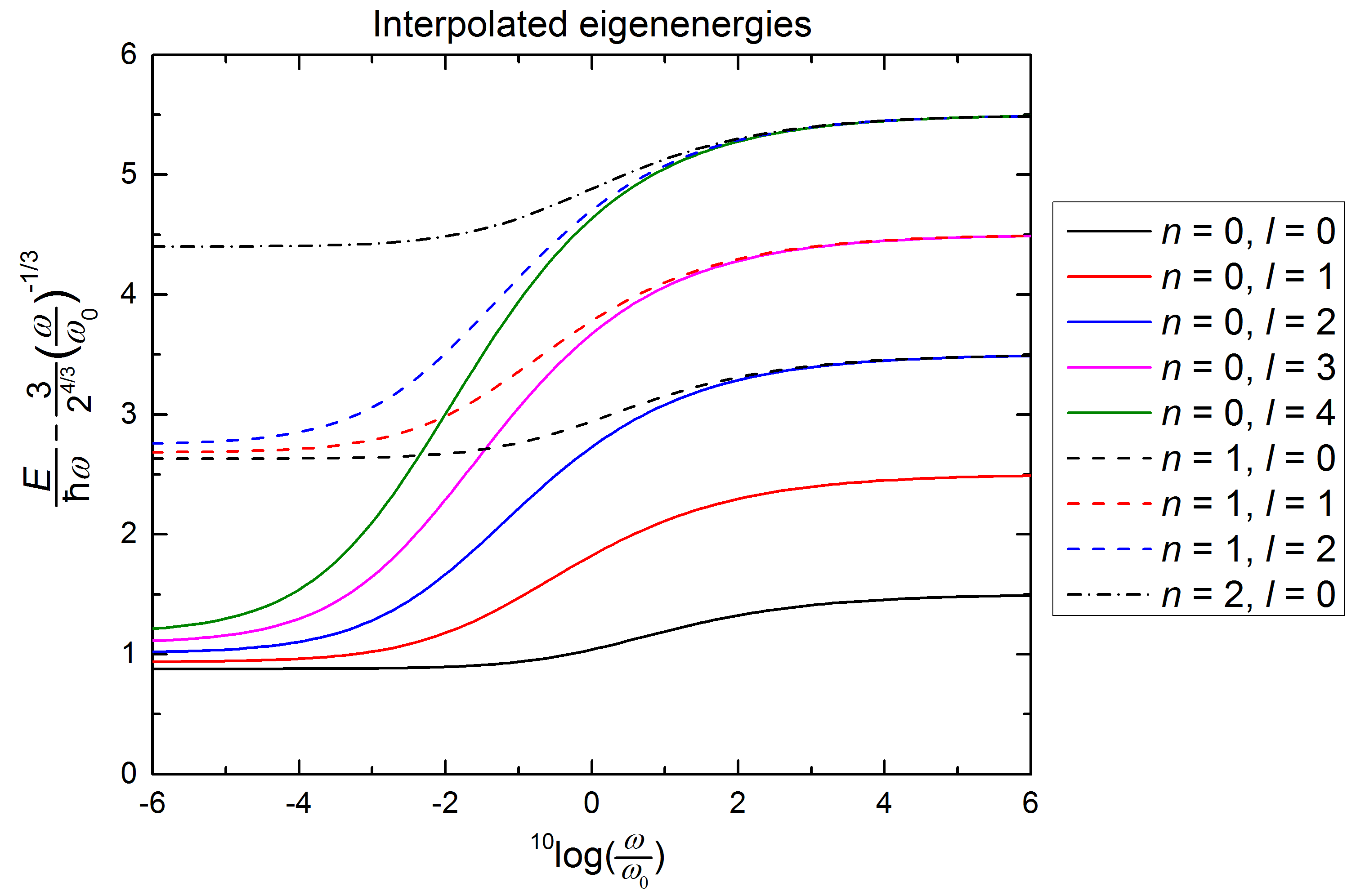}
\end{center}
\vspace{-0.75cm}
\caption{\textrm{Eigenenergies for a number of states, with quantum numbers as indicated. The asymptotic behavior for $\omega\rightarrow0$ has been subtracted for a clearer view on the results.}}
\label{General_states}
\end{figure}

Fitting \refEquation{HF_diff_better} through the exact solutions found by Taut's method we can determine the eigenenergies as function of the frequency for general quantum numbers $n$ and $l$. The found energies are presented in \refFigure{General_states}, where the asymptotic behavior for small frequencies given by the first term of \refEquation{Energy_low_w} is subtracted for clarity.

The figure might indicate that for $\omega\rightarrow0$ the energy does not depend on the quantum number $l$. With the full effective potential of \refEquation{Effective_pot}, the harmonic approximation \refEquation{HO_low_w}, and expanding the resulting energy up to first order in $l(l+1)$ we find
\begin{equation}
\frac{E}{\hbar\omega}=\frac{E_{l=0}}{\hbar\omega}+\frac{E^{(1)}(\omega)}{\hbar\omega}l(l+1),
\end{equation}
where $E_{l=0}$ is given by \refEquation{Energy_low_w} and
\begin{equation}
\frac{E^{(1)}(\omega)}{\hbar\omega}=\frac{1}{2^{2/3}}\left(\frac{\omega}{\omega_0}\right)^{1/3}+\frac{1}{2^{1/3}\sqrt{3}}\left(n+\frac{1}{2}\right)\left(\frac{\omega}{\omega_0}\right)^{2/3}.
\end{equation}
This result was found numerically in the context of harmonium \cite{Karwowski04}, but here we derived it analytically, which allows for a scaling in terms of the Hartree frequency. Higher order corrections can also be found analytically. Now, since $\lim\limits_{\omega\rightarrow0} \frac{E^{(1)}(\omega)}{E_{l=0}}=0$, the energy does not depend any more on $l$ in the limit of low frequencies, and is simply given by \refEquation{Energy_low_w}. This was to be expected since we already had established that the importance of the Coulomb term in \refEquation{Effective_pot} is negligible, and therefore the centrifugal barrier term should be negligible as well. The limiting case in which the  energy is independent of the quantum number $l$, indicates that states with different angular momenta are equally important. This is in agreement with the partial wave treatment of Coulomb scattering \cite{sakurai}, which formally corresponds to $\omega=0$.

\section{Many-particle systems}
A gas of degenerate electrons as mentioned in the introduction, is generally a multi-electron system. If we assume only pairwise interactions between electrons, we can easily use the results from previous sections to find the ground state by populating the two-particle states given in \refFigure{General_states}. In the weakly-coupled regime this corresponds to filling the eigenstates of a simple harmonic oscillator from below. In the strongly-coupled regime, for sufficiently small $\omega$, all pairs are in states with $n=0$. In the limit of  $\omega\rightarrow0$, these states all become energetically degenerate which is of course a non-trivial situation for fermions.

With the techniques shown in this paper, finding analytical expressions for the energies of eigenstates in a many-particle electron system should be possible. In the weakly-coupled regime the Coulomb interaction energy is much smaller than the harmonic oscillator energy separation. A many-particle electron system behaves then as a weakly-interacting Fermi gas with the pairwise Coulomb interaction energy as a perturbation. Such an electron system can be quantum degenerate if its temperature is below the Fermi temperature of the harmonic oscillator. In the strongly-coupled regime where the Coulomb interaction energy dominates, perturbative methods seem at first sight not applicable. However, subtraction of the diverging $\omega^{2/3}$ energy scaling, \refEquation{scalinglaw}, results in a harmonic oscillator problem with a factor $\sqrt 3$ frequency and mixed angular momentum states, where the remaining Coulomb contribution is again small compared to the harmonic oscillator energy separation. This, combined with some of the results in the context of harmonium, suggests that also here a perturbative many-body approach is possible.

In fact, for $N$ trapped electrons it can be shown \cite{CioslowskiPRA} that the Hamiltonian is given by
\begin{equation}
H=-\frac{\hbar^2}{2m_e}\sum\limits_{p=1}^N\nabla_p^2+\frac{1}{2}m_eN\omega^2R^2+\sum\limits_{p>q=1}^N\left[\frac{m_e\omega^2}{2N}r_{pq}^2+\frac{k}{r_{pq}}\right],
\end{equation}
where $R=\frac{1}{N}\sum\limits_{p=1}^Nr_p$, $r_{pq}=r_p-r_q$ and $\nabla_p=\frac{\partial}{\partial r_p}$ with $r_p$ the position of the $p$-th electron. In the weakly-correlated regime, we neglect the Coulomb repulsion like we did for $N=2$, so that the potential is purely harmonic, and therefore the relative energies are given by
\begin{equation}\label{EN_high_w}
\frac{E}{\hbar\omega}=\sum\limits_{i=1}^{N-1}\left(2n_i+l_i+\frac{3}{2}\right),
\end{equation}
which indeed describes filling a simple harmonic oscillator with electrons from below. An approximation for the energy in the strongly-correlated limit can easily be found if we assume that the coordinates $r_{pq}$ in the Hamiltonian are independent. Then, making a harmonic approximation around the minimum of each of the pair-potentials and by taking into account the multiplicity of the double sum, we can solve for this Hamiltonian in terms of $\frac{1}{2}N(N-1)$ single-particle harmonic oscillator states with a total energy given by
\begin{align}\label{EN_low_w}
  \begin{split}
  \frac{E}{\hbar\omega} &= \frac{3}{4}N^{2/3}(N-1)\left(\frac{\omega}{\omega_0}\right)^{-1/3}\\
    &\hspace{3em} +\sum\limits_{i=1}^{\frac{1}{2}N(N-1)}\left(n_i+\frac{1}{2}\right)\sqrt{3}
  \end{split}\\
  &\equiv A_N\left(\frac{\omega}{\omega_0}\right)^{-1/3}+B_N+\sqrt{3}\sum\limits_{i=1}^{\frac{1}{2}N(N-1)}n_i.
\end{align}
The first term in \refEquation{EN_low_w} corresponds to a frozen-gas approximation where the kinetic energy of all electrons is neglected and simply the minimum of each pair-potential is determined. The minimum of the potential results in an equilibrium position between the electrons, where the spatial positions form a crystal-like pattern. The corresponding distance that minimizes the pair-potential is $r_0=a_0N^{1/3}(\omega/\omega_0)^{-2/3}$ for any $N$. The electron gas in this regime is equivalent to a Wigner crystal \cite{wigner1934interaction,Cioslowski06,Cioslowski12b} up to $N=4$. The resulting coefficient $A_N=3N^{2/3}(N-1)/4$ matches exactly the expressions found in \refEquation{Energy_low_w} for $N=2$ and in Refs.~\cite{Cioslowski06,CioslowskiPRA} for $N=3$ and $N=4$. The second term in \refEquation{EN_low_w} corresponds to vibrational states of the crystal. Comparing the coefficient $B_N=N(N-1)\sqrt{3}/4$ to the results for $N=2$ in \refEquation{Energy_low_w} and $N=3,4$ in Refs.~\cite{Cioslowski06,CioslowskiPRA} we see that the relative difference is 0\%, 24\% and 46\% respectively. We interpret this difference as correlation energy due to the angular correlations between vibrational states of the crystal, which are correctly accounted for in Refs.~\cite{Cioslowski06,CioslowskiPRA} but we neglected by assuming that all $r_{pq}$ are independent. However, far into the strongly correlated regime the resulting energy corrections are relatively small compared to the total energy $E$. Thus we see that also for $N$ electrons, harmonic approximations in both the weakly and strongly correlated regime indeed are suitable methods to find the eigenenergies and corresponding wave functions.

The Hellman-Feynman interpolation method that we presented can also be generalized to many-particle electron systems. However, in contrast to $N=2$ where we used Taut's exact solutions to determine the fitting parameters $\alpha$ and $\beta$ in \refEquation{HF_diff_better}, we would have to resort to numerical solutions as in Refs.~\cite{Cioslowski12b,Cioslowski14}. Furthermore, to find the generalized form of \refEquation{HF_diff_better}, we can use the expressions \refEquation{EN_high_w} and \refEquation{EN_low_w}, or if a higher accuracy is needed one could use the results from Refs.~\cite{Cioslowski06,CioslowskiPRA}.

\section{Conclusion}
We described the problem of two electrons trapped in a harmonic potential for any value of the trapping frequency. Two different regimes can be identified: the strongly-coupled regime for small frequencies, and the weakly-coupled regime for large frequencies. In both regimes we found a harmonic approximation of the effective potential to correctly produce the eigenenergies of the electron pair for the ground state as well as any excited state. The critical crossover trapping frequency corresponds to the Hartree frequency. Using the Hellmann-Feynman theorem we were able to find semi-analytical solutions to the problem at any given frequency and for any given state. As shown, these concepts are readily generalized to many-particle systems.

The trapping frequencies for current electron traps are several orders of magnitude below the crossover frequency. This implies that present day experiments are far into the strongly-coupled limit. Conversely, in the context of harmonium trapping frequencies are typically above the crossover frequency \cite{CioPer}. Interestingly, as we demonstrated in this letter, both these regimes can be described analytically, experimentally relevant to the experiments proposed on entangled electron pairs \cite{gabrielse_two_electron,Lougovski11}. Moreover, our methodology allows for a perturbative many-body calculation based on a paired-electron description, as well as a feasible interpolation method. This allows us to explore the properties and phases of many-particle quantum-degenerate electron systems under harmonic confinement.


\begin{acknowledgments}
This research was financially supported by the Foundation for Fundamental Research on Matter (FOM).
\end{acknowledgments}

\bibliographystyle{apsrev}
\bibliography{Bibliography}

%

\end{document}